\documentclass[prd,twocolumn,amsmath,amssymb,floatfix,superscriptaddress,nofootinbib]{revtex4-1}

\usepackage{graphicx}
\usepackage{amssymb}
\usepackage{amsmath}
\usepackage{color}

\DeclareMathAlphabet\mathbfcal{OMS}{cmsy}{b}{n}
\definecolor{darkgreen}{cmyk}{0.85,0.2,1.00,0.35}
\definecolor{purple}{cmyk}{0.5,1.0,0,0}

\newcommand{\stucky}{St\"{u}ckelberg}
\newcommand{\Gs}{\delta \Gamma}
\newcommand{\tr}[1]{[#1]}
\newcommand{\dd}{{\rm d}}
\newcommand{\ul}[3]{#1^{#2}_{ \hphantom{#2}\! #3}}

\newcommand{\bgamma}{\boldsymbol{\gamma}}
\def\be{\begin{equation}}
\def\ee{\end{equation}}
\def\ben{\begin{equation} \nonumber}
\def\een{\end{equation}}
\def\ban{\begin{eqnarray*}}
\def\ean{\end{eqnarray*}}
\def\ba{\begin{eqnarray}}
\def\ea{\end{eqnarray}}
\def\({\left(}
\def\){\right)}

\newcommand{\Comment}[1]{{}}
\definecolor{darkblue}{rgb}{0.15,0.35,0.55}
\definecolor{reddish}{rgb}{0.65, 0.2, 0.2}
\definecolor{darkgreen}{RGB}{50,150,0}
\usepackage[linktocpage=true]{hyperref}
\hypersetup{
colorlinks=true,
citecolor=darkblue,
linkcolor=reddish,
urlcolor=darkblue,
pdfauthor={},
pdftitle={},
pdfsubject={}
}

\begin{document}

\title{Self-accelerating Massive Gravity: \\  Superluminality, Cauchy Surfaces and Strong Coupling}
\author{Pavel Motloch}
\affiliation{Kavli Institute for Cosmological Physics, Department of Physics, University of Chicago, Chicago, Illinois 60637, U.S.A}
\author{Wayne Hu}
\affiliation{Kavli Institute for Cosmological Physics, Department of Astronomy \& Astrophysics,  Enrico Fermi Institute, University of Chicago, Chicago, Illinois 60637, U.S.A}

\author{Austin Joyce}
\affiliation{Kavli Institute for Cosmological Physics, Department of Astronomy \& Astrophysics,  Enrico Fermi Institute, University of Chicago, Chicago, Illinois 60637, U.S.A}

\author{Hayato Motohashi}
\affiliation{Kavli Institute for Cosmological Physics, Department of Astronomy \& Astrophysics,  Enrico Fermi Institute, University of Chicago, Chicago, Illinois 60637, U.S.A}

\begin{abstract}
\noindent
Self-accelerating solutions in massive gravity provide explicit, calculable
examples that exhibit the general interplay between superluminality, the well-posedness of
the Cauchy problem, and strong coupling. For three {particular} classes of vacuum solutions, one of which is new to this work, we construct the conformal diagram for the characteristic surfaces on which isotropic stress-energy perturbations propagate. With one exception, all solutions necessarily possess spacelike characteristics, indicating perturbative superluminality.
Foliating the spacetime with these surfaces gives a pathological frame where kinetic terms of the perturbations vanish, confusing the Hamiltonian counting of degrees of freedom. This frame dependence distinguishes the vanishing of  kinetic terms from strong coupling of perturbations or an ill-posed Cauchy problem. We give examples where spacelike characteristics do and do
not originate from a point where perturbation theory breaks down and where  spacelike  surfaces do or do not intersect all characteristics in the past light cone
of a given observer. The global structure of  spacetime also reveals issues that are unique to theories with two metrics: in all three classes of solutions, the Minkowski fiducial space fails to cover the entire de Sitter spacetime
allowing worldlines of observers to end in finite proper time at determinant singularities. Characteristics run
tangent to these surfaces requiring {\it ad hoc} rules to establish continuity across 
singularities.
\end{abstract}

\maketitle

\section{Introduction}

Building upon earlier work~\cite{ArkaniHamed:2002sp,Creminelli:2005qk,deRham:2010ik}, 
de Rham, Gabadadze and Tolley (dRGT) \cite{deRham:2010kj} first constructed a consistent interacting theory of a massive spin-2 field using an auxiliary flat fiducial metric.  Though the study of this theory and other nonlinear massive gravity variants has now reached a somewhat mature stage \cite{deRham:2014zqa},  there remain many outstanding questions to be addressed and understood.

Foremost amongst these is the status of cosmological solutions within the theory. 
The dRGT theory does not admit
flat or closed cosmological solutions where the spacetime and fiducial metrics are simultaneously homogeneous and isotropic~\cite{D'Amico:2011jj,Gumrukcuoglu:2011ew}. Though this complicates matters computationally, it does not 
automatically destroy the phenomenological viability of cosmological solutions.  The spacetime
upon which matter fields propagate can be homogeneous and isotropic for any choice of spatial curvature
\cite{Gratia:2012wt,Volkov:2012cf,Volkov:2012zb}.   Indeed, many explicit examples of cosmological solutions to the theory have been found where accelerated expansion occurs without a true cosmological
constant~\cite{deRham:2010tw,Koyama:2011xz,Gumrukcuoglu:2011ew,Koyama:2011yg,Nieuwenhuizen:2011sq,
Berezhiani:2011mt,D'Amico:2011jj,Gratia:2012wt,Kobayashi:2012fz,
Volkov:2012cf,Volkov:2012zb}.

Though these solutions are satisfactory as far as background evolution is concerned, the theory of fluctuations about these solutions is somewhat paradoxical.  Analyses of perturbations about these self-accelerating backgrounds~\cite{Wyman:2012iw,Khosravi:2013axa} reveal pathologies such as unboundedness of the Hamiltonian for isotropic perturbations, or strong-coupling of degrees of freedom in specific backgrounds, but often these appear in quite a subtle way \cite{Motloch:2014nwa}.  For example it has been claimed
that the number of degrees of freedom identified by a Hamiltonian analysis depends
on a choice of coordinates \cite{Khosravi:2013axa}.  Therefore, a worthwhile aim is to understand more fully the relationship between these potentially problematic features.

More generally, this broad class of self-accelerating solutions provides an interesting playground to explore fundamental field-theoretical questions which remain unresolved in massive gravity and beyond. In particular, there has been much recent interest in the interplay between superluminality in field theories and strong coupling phenomena, as well as what if anything the presence of these features implies about the well-posedness 
of the Cauchy problem (for various perspectives, see~\cite{Adams:2006sv,Dubovsky:2007ac,Shore:2007um,deRham:2014lqa,Hollowood:2009tw,Hinterbichler:2009kq,Porrati:2009bs,Goon:2010xh,Padilla:2010tj,deFromont:2013iwa,Cooper:2013ffa,Joyce:2014kja,deRham:2013hsa,Creminelli:2014zxa,Keltner:2015xda,Deser:2012qx,Deser:2013eua,Deser:2014hga,Deser:2014fta,Deser:2015wta}). The self-accelerating sector of dRGT allows us an opportunity to explore these issues with explicit solutions: the theory of isotropic perturbations about these solutions turns out to be rather simple \cite{Wyman:2012iw}. In particular, it is possible to solve exactly for the characteristic hypersurfaces upon which graviton stress energy fluctuations propagate. This makes it rather easy to study perturbative aspects of the causal structure of self-accelerating backgrounds.
Therefore, another worthwhile avenue to explore is to what extent we can abstract general lessons about superluminality, strong coupling and Cauchy surfaces from these particular examples.

In this paper, we pursue both of these goals in tandem. After reviewing the structure of the dRGT theory---and in particular the construction of self-accelerating solutions of Ref.~\cite{Gratia:2012wt}---we focus on three particular families of vacuum solutions, one of which is new
to this work.  We explicitly solve for the characteristics for isotropic perturbations and examine their causal structure
via conformal diagrams.   In this sense, our study is related to the analysis of  Ref.~\cite{Deser:2012qx,Izumi:2013poa,Deser:2013eua,Deser:2014hga,Deser:2014fta,Deser:2015wta} but our exact solutions are general and do not require discontinuous or ``shock" conditions which automatically entail strong coupling.  We find many interesting phenomena: in particular,  the solutions we consider exhibit  superluminal propagation of fluctuations  by necessity except in one unique case, only some of which provide a well-posed Cauchy problem and originate from singular  conditions where strong-coupling might reside.

Perturbative superluminality on specific backgrounds does not necessarily indicate superluminality in the full theory
(see \cite{Dubovsky:2007ac,Shore:2007um,deRham:2014lqa}), which would present problems for
a local and Lorentz-invariant UV completion of the theory itself~\cite{Adams:2006sv},
nor does it imply
acausal structures such as closed timelike curves \cite{Babichev:2007dw,Burrage:2011cr}.
However these examples do highlight the fact that the highly related notions of superluminality, strong coupling and well-posed Cauchy problem are conceptually distinct.

Our examples also highlight an issue that is unique  to a theory with two metrics.
In all three families of vacuum solutions, the Minkowski fiducial space fails to cover the
entire spacetime.   The point at which the Minkowski chart, or unitary gauge, ends is
called a determinant singularity; it is diffeomorphism-invariant
and hence physical \cite{Gratia:2013gka,Gratia:2013uza}.    In these examples, worldlines of some observers
can intersect the singularity in finite proper time and for others the singularity lies
in their past light cone requiring {\it ad hoc} rules for continuing worldlines or boundary conditions.

The paper is organized as follows.  
In \S \ref{sec:theory}, we review the construction of bi-isotropic self-accelerating background solutions
\cite{Gratia:2012wt} and isotropic perturbations around them \cite{Wyman:2012iw}.  We then
specialize to the  vacuum case in  \S \ref{sec:vacuum} and discuss the global structure of the spacetime background and fiducial metric for three families
of solutions, one of which is new to this paper.
In \S \ref{sec:perturb}, we employ a  characteristic analysis of perturbations around these backgrounds to
expose the relationship between superluminality, strong coupling and the well-posedness of the 
Cauchy problem.
We discuss the implications of these results in \S \ref{sec:discuss}.

\section{Self-Acceleration in Massive Gravity}
\label{sec:theory}

In this section we briefly review the properties of the dRGT theory of massive gravity, the construction of self-accelerating background solutions, and spherically symmetric perturbations around them.

\subsection{Fiducial Metric}

The dRGT \cite{deRham:2010kj} nonlinear theory of a massive spin-2 field
is given by the following action which propagates only the expected 5 polarizations {of a massive graviton}: 
\be
\label{drgt}
S = \frac{M_{\rm Pl}^2}{2}\int\dd^4x\sqrt{-g}\left(R-m^2\sum_{k=0}^4 \frac{\beta_k}{k!} F_k\left(\bgamma\right)
\right),
\ee
where $M_{\rm Pl}^2 = (8\pi G)^{-1}$ is the reduced Planck mass
and the $F_k$ terms are
characteristic polynomials of the matrix $\bgamma$. These can be written explicitly as
\begin{align}
F_0(\bgamma) & = 1, \nonumber\\
F_1(\bgamma) & = \tr{\bgamma}, \nonumber\\
F_2(\bgamma) & =  \tr{\bgamma}^2 - \tr{\bgamma^2} , \\
F_3(\bgamma) & =\tr{\bgamma}^3 - 3 \tr{\bgamma} \tr{\bgamma^2} + 2 \tr{\bgamma^3} , \nonumber\\
F_4(\bgamma) &= \tr{\bgamma}^4 - 6 \tr{\bgamma}^2 \tr{\bgamma^2} + 3 \tr{\bgamma^2}^2 + 8 \tr{\bgamma} \tr{\bgamma^3}
- 6 \tr{\bgamma^4} ,
\nonumber
\end{align}
where $[\,]$ denotes the trace of the enclosed matrix. The matrix $\bgamma$ is the  square root of the  product of the inverse spacetime metric ${\bf g}^{-1}$ and a flat fiducial metric $\boldsymbol{\Sigma}$
\be
\ul{\gamma}{\mu}{\alpha} \ul{\gamma}{\alpha}{\nu} = g^{\mu\alpha}\Sigma_{\alpha\nu} \,.
\label{eqn:gamma}
\ee

The parameters of the theory defined by Eq.~\eqref{drgt} are
$m$---the graviton mass---and the $\beta_k$.  These parameters are not all independent, but rather they depend on two fundamental independent parameters~$\{\alpha_3,\alpha_4\}$ through
\begin{align}
\beta_0 &= -12 (1+ 2\alpha_3+2\alpha_4), \nonumber\\
\beta_1 &= 6(1 + 3 \alpha_3 + 4\alpha_4),\nonumber\\
\beta_2 &= -2(1+ 6 \alpha_3+12\alpha_4 ), \\
\beta_3 &= 6(\alpha_3+ 4\alpha_4), \nonumber\\
\beta_4 &= -24 \alpha_4.\nonumber
\end{align}

The chart of the spacetime metric for which the fiducial metric is represented by the Minkowski metric 
$\Sigma_{\mu\nu}=\eta_{\mu\nu}$ is called
unitary gauge and there the spacetime metric possesses additional degrees of freedom compared to Einstein gravity.
Diffeomorphism invariance can be restored by employing the St\"uckelberg trick to introduce four scalar functions, $\phi^a$, to represent the flat fiducial metric in arbitrary coordinates
\be
 \Sigma_{\mu\nu} = \eta_{a b} \partial_\mu\phi^a\partial_\nu\phi^b .
\ee
The $\phi^a$ are equal to the unitary gauge coordinates $x_u^\mu$ and  absorb the  extra polarization states in arbitrary gauges where diffeomorphism invariance is used to eliminate or constrain
 the metric terms. 
The matrix $\partial_\mu\phi^a$ then represents the Jacobian of the coordinate transform between a general set of coordinates
$x^\mu$ and $x_u^\mu$.

Minkowski space may not cover, or equivalently unitary gauge may not chart, the entire spacetime; this situation is signaled by  a non-invertible Jacobian
transform, which we call a determinant singularity \cite{Gratia:2013gka}. 
Unlike a pure coordinate singularity, a determinant singularity  does not depend on the
chart of the spacetime.   This is because the ratio of determinants of the two metrics 
\begin{equation}
{\rm det}({\bf g}^{-1} {\boldsymbol{\Sigma}})={\rm det}({\bf g}_u^{-1} {\boldsymbol{\eta}})
=-{\rm det}({\bf g}_u^{-1})
\end{equation}
is a spacetime scalar.  A coordinate singularity in the unitary chart of the spacetime metric ${\bf g}_u$ therefore becomes a
coordinate-invariant determinant singularity. In contrast to a curvature singularity, the two metrics  need not individually have any diffeomorphism invariant singularities, despite the fact that combined they display a determinant singularity.

Physically, the presence of two metrics means that worldlines
of observers may end after a finite proper time in spacetime has elapsed, since 
an infinite interval can elapse as measured by the fiducial metric. This geodesic incompleteness is another indicator that we should take determinant singularities seriously.
It is possible to continue worldlines past these singularities in the spacetime, but this requires multiple copies of  Minkowski space and a rule for
continuing the chart---or equivalently the \stucky\ fields---that is not directly imposed by the action.

\subsection{Background Equations of Motion}
\label{sec:backgroundeoms}

For any isotropic spacetime metric,
including cosmological solutions with arbitrary matter content, there are solutions to the dRGT equations of motion
where the stress-energy associated with the graviton potential in Eq.~\eqref{drgt} behaves as a cosmological constant. 
The construction of Ref.~\cite{Gratia:2012wt}, which we now review, is in fact extensible beyond dRGT to cases with non-flat bi-isotropic and dynamical metrics \cite{Motohashi:2012jd,Gratia:2013uza}.

Any metric with {spatial slices invariant under SO(3) rotations} can be written in isotropic coordinates, in which the line element takes the form
\be
	{\dd}s^2 = -b^2(t,r) \dd t^2 + a^2(t,r) \big(\dd r^2 + r^2  \dd \Omega_2^2 	\big),
	\label{isotropicmetric}
\ee
where $\dd \Omega_2^2$ is the line element on a 2-sphere. If the fiducial metric is rotationally invariant
in the same coordinate system ({\it i.e.}, the metrics are bi-isotropic), 
the \stucky\ fields take the following form,
\ba
	 \phi^0 &=& f(t,r),\nonumber\\
	 \phi^i &=& g(t,r) \frac{x^i}{r},
	\label{stuckyback}
\ea
and are completely specified by the two functions $f$ and $g$.
Note that $f =  t_u$ is the unitary gauge time  and $g=r_u$ is the unitary gauge radius that describe the
fiducial flat line element
\begin{eqnarray}
\dd s_\Sigma^2 &=& \Sigma_{\mu\nu} \dd x^\mu \dd x^\nu
= -\dd f^2 + \dd g^2 + g^2 \dd \Omega_2^2 .
\label{eqn:fidline}
\end{eqnarray}
Unitary gauge uniquely
specifies the coordinates, whereas the isotropic condition does not, leading to multiple paths to
finding the same solution and superficially different descriptions of their dynamics.
Once solutions for $f$ and $g$ are found using any isotropic construction they may be 
re-expressed in an alternate choice of coordinates since both are spacetime scalars.

Upon inserting the ans\"atze~\eqref{isotropicmetric} and~\eqref{stuckyback} into the  action~\eqref{drgt}, we find that the equation of motion for the spatial St\"uckelberg fields is satisfied by
\ba
	g(t,r) =x_0  a(t,r)r,
	\label{eqn:gsoln}
\ea
where the constant $x_0$ solves the polynomial equation $P_1(x_0)=0$ with
\be
P_1(x) \equiv 2(3-2x)+6(x-1)(x-3)\alpha_3+24(x-1)^2\alpha_4.
\ee
On this solution, the effective stress tensor due to the presence of the non-derivative graviton interactions takes the form of an effective cosmological constant
\be
T_{\mu\nu} = - \Lambda_{\rm eff} M_{\rm Pl}^2 g_{\mu\nu},
\ee 
where
\be
\Lambda_{\rm eff}=  \frac{1}{2} m^2 P_0(x_0),
\ee
and the polynomial $P_0(x)$ is given by
\begin{align}
P_0(x) &= - 12 - 2 x(x-6) - 12(x-1)(x-2)\alpha_3 
\nonumber\\&\qquad -24(x-1)^2\alpha_4 .
\end{align}
Unitary time, $f(t,r)$, satisfies the equation
\begin{equation}
\sqrt{X} = \frac{W}{x_0}+x_0,
\label{eqn:feqnsoln}
\end{equation}
where \begin{align}
X & \equiv\Bigl(\frac{\dot{f}}{b}+\mu\frac{g'}{a}\Bigr)^2-\Bigl(\frac{\dot{g}}{b}+\mu\frac{f'}{a}\Bigr)^2, \nonumber\\
W & \equiv \frac{\mu}{ab} \( \dot f g' - \dot g f' \),
\label{eqn:XW}
\end{align}
with branches due to the matrix square root $\bgamma$ defined in Eq.~\eqref{eqn:gamma} allowing
$\mu\equiv \pm 1$.  Here and throughout, overdots denote derivatives with respect to $t$ and primes denote
derivatives with respect to $r$.
Note that $W$ is proportional to the determinant
of the Jacobian transform from unitary gauge to isotropic coordinates. When $W=\pm\infty,
0$ or is undefined 
because either $f$ or $g$
are not continuously differentiable, the Jacobian
transform is not invertible.   We call all of these cases a determinant singularity.   Analytic continuation is sometimes possible, especially in the later two cases, but requires a second fiducial metric and solution with its own choice
of branch  \cite{Gratia:2013gka}. We pick $\mu=1$ for the examples in the following sections.

Using Eq.~\eqref{eqn:gsoln} in Eq.~\eqref{eqn:feqnsoln}, we see that the latter equation can be cast as
\ba
\label{fEOMback}
	b^2 f'^2 + 2 a r(a' \dot f^2 - \dot a \dot f f')+r^2 (a' \dot f - \dot a f')^2 \nonumber
	\\
	= x_0^2 \(a'^2 b^2 r^2 + 2 a' a b^2 r - \dot a^2 a^2 r^2\) .
\ea
This nonlinear partial differential equation has an infinite number of distinct self-accelerating solutions, each of
which possesses the same
background spacetime metric and $\Lambda_{\rm eff}$. In order to specify a solution, 
one assumes a functional form for $a$ and $b$ consistent with the Einstein equations sourced by
$\Lambda_{\rm eff}$ and the matter content 
and then solves Eq.~\eqref{fEOMback} to determine $f$.  In \S \ref{sec:vacuum}, we will perform this procedure, choosing $a$ and $b$ so that there is no matter content and the background spacetime is de Sitter.

For solutions to the equation of motion~\eqref{fEOMback} which satisfy the condition
\begin{equation}
\dot f f' = \dot g g',
\label{eqn:bidiagonal}
\end{equation}
the fiducial metric Eq.~\eqref{eqn:fidline} is also diagonal in the same $(t,r)$ coordinates that the spacetime
metric is diagonal.   We shall see that this is 
exactly the condition for which the kinetic term of isotropic perturbations vanishes  \cite{Khosravi:2013axa}.
However, we can analyze the dynamics  of the same solution in alternate isotropic coordinate systems which mix the temporal and radial coordinates where bi-diagonality and vanishing kinetic terms do not apply. 
Moreover, 
the diffeomorphism invariance of the \stucky\ representation means that we are not even limited to
isotropic coordinate systems when analyzing the dynamics.

\subsection{Isotropic Perturbations}
\label{sec:isopert}

Given the rotational invariance of the background spacetime metric and St\"uckelberg fields, isotropic perturbations about these background solutions are straightforward  to analyze ({\it cf.}\ \cite{Motloch:2014nwa}). This is a reasonable starting point, as uncovering  problems
in this restricted class of perturbations would already indicate a pathology of the background solution.
Note that the converse is not true: a
background solution with healthy isotropic perturbations can still show pathologies in the
anisotropic sector.

In order to study fluctuations about the background solutions we are considering, we perturb both the metric variables and St\"uckelberg fields about their background values; 
isotropic perturbations are specified by four functions of $(t,r)$: 
$\delta a, \delta b, \delta f$ and $\delta g$.
Varying the quadratic action for these variables given in Ref.~\cite{Wyman:2012iw} yields 
an independent equation of motion for a specific combination
\be
	\delta \Gamma(t,r) = \delta g(t,r) - x_0  r \delta a(t,r) .
\ee
This combination is special because it is precisely the variable which quantifies perturbations away from the effective cosmological constant solution; perturbations for which $\delta \Gamma = 0$ still satisfy~\eqref{eqn:gsoln} and therefore 
 leave $\Lambda_{\rm eff}$  unchanged. The equation of motion for the variable $\delta\Gamma$ is~\cite{Wyman:2012iw}
\begin{eqnarray}
&& \partial_t \left[ 
\frac{ a^2 r }{\sqrt{X}} \left( \frac{\dot f}{b} + \mu \frac{g'}{a} \right)\Gs \right] =
\partial_r \left[
\frac{ab r }{\sqrt{X}} \left( \mu \frac{\dot g}{b} + \frac{f'}{a}\right)\Gs\right] 
\nonumber\\
&&\qquad + \mu a^2 r^2  \left[ \frac{(a r)'}{ar } \dot \Gs -  \frac{\dot a}{a}\Gs' \right] .
\label{eqn:gs}
\end{eqnarray}
Equation~\eqref{eqn:gs} is first order in both time and space derivatives and does not depend on
other perturbations. The $\delta f$ equation of motion is also first order, but
unlike the $\Gs$ equation it requires specification of $\Gs$ and other perturbations.
To linear order in the \stucky\ fluctuations, the stress energy tensor of the perturbations depends only
on $\Gs$ and we therefore focus on its dynamics.  The equation of motion \eqref{eqn:gs} is equivalent to local conservation of stress energy \cite{Wyman:2012iw}.

The coefficient of the temporal derivative term,
\begin{eqnarray}
\label{At}
A_t = \frac{ a^2 r }{\sqrt{X}} \left( \frac{\dot f}{b} + \mu \frac{g'}{a} \right) - \mu a r {(a r)'},
\end{eqnarray}
is of special interest to the time evolution of the perturbations.  
In a choice of isotropic coordinates
where $A_t=0$, the kinetic term for $\delta \Gamma$ vanishes and hence the equation of 
motion \eqref{eqn:gs} does not specify the evolution of perturbations off of a constant time surface. 
Combining Eq.~\eqref{fEOMback} and Eq.~\eqref{eqn:bidiagonal}, we see that $A_t=0$ whenever the spacetime and
fiducial metrics are bi-diagonal and vice versa.

When $A_t=0$, the energy density, momentum and pressure associated with any $\Gs$ fluctuation vanish to 
linear order
\cite{Wyman:2012iw}.  Thus energy-momentum conservation, though enforced by the equation of motion,
also does not yield equations that allow evolution of $\Gs$ off of constant time surfaces. 
Instead, the nonvanishing anisotropic stresses must obey a constraint equation.

On the other hand,
the fact that the equation of motion 
can be written in terms of different choices of isotropic time---or more generally
different foliations of the spacetime---indicates that the
vanishing of the kinetic term is a coordinate dependent statement~\cite{Khosravi:2013axa}.\footnote{Note that there are more prosaic examples of this phenomenon: a scalar field with a superluminal phase velocity considered in an appropriately Lorentz-boosted frame will have a vanishing kinetic term \cite{Adams:2006sv}.}
In a choice where the kinetic term is present, the equation of motion does supply 
an evolution equation, or equivalently the constraint
on stresses becomes a non-trivial equation for energy-momentum conservation.
We shall see that this is a generic property for perturbations that propagate superluminally, {\it i.e.}, on spacelike characteristics.

Relatedly, Ref.~\cite{Khosravi:2013axa} show through a Hamiltonian analysis
that the number of propagating (isotropic) degrees of freedom in a given
isotropic frame is one if $A_t \neq 0$ and zero if $A_t =0$.   The number of
degrees of freedom counted in this way, therefore depends on the coordinate
system.   For $A_t = 0$, the Hamiltonian is not associated with the time evolution of the system, in the sense that it defines evolution along a spatial slice, rather than transverse to it. As such, it is somewhat unclear how to interpret the Hamiltonian analysis in this case.
Further, in the case $A_t \neq 0$, the canonical momentum for $\delta \Gamma$ appears linearly in the Hamiltonian, and thus the Hamiltonian is unbounded from below.

\begin{figure*}
\center
\includegraphics[width = 0.99\textwidth]{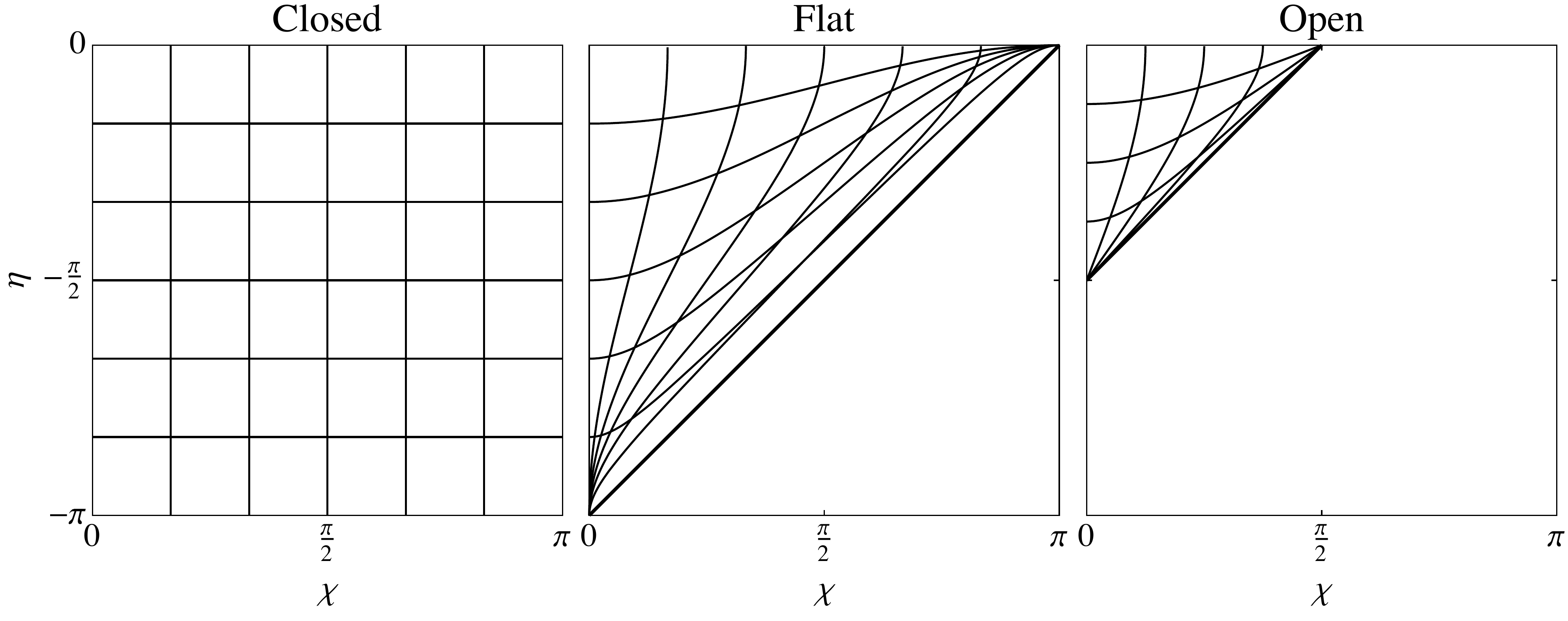}
\caption{Conformal diagrams showing portions of de Sitter space charted by closed (left),
flat (middle) and open (right) foliations. Thick  lines indicate coordinate
singularities. Superimposed are lines of constant isotropic time and radius for each foliation.
 }
\label{fig:coverings}
\end{figure*}

\section{Vacuum Solutions}
\label{sec:vacuum}

In this section we specialize to background solutions without any matter content. In these
solutions the only source of stress-energy determining the background is the effective
cosmological constant, $\Lambda_{\rm eff}$, and the spacetime metric describes a de Sitter
space.  We first discuss the conformal and three isotropic charts of the de Sitter space that
play a prominent role in both the construction of solutions and the investigation of their global properties.  
We then turn to three families of self-accelerating solutions and discuss their global structure, namely the appearance of their determinant singularities.

\subsection{De Sitter Charts}

One special feature of de Sitter space is that there is no preferred temporal coordinate to define a foliation with respect to. Sections of the full spacetime can therefore be charted by isotropic
coordinates where the constant time slices have positive, negative or zero spatial curvature.
The conformal diagram for de Sitter space can be constructed from the  positive curvature (closed) foliation of the spacetime, where the line element takes the form
\begin{equation}
\label{conformal_metric}
\dd s^2 = \left( \frac{1}{H \sin \eta}\right)^2 \left(-\dd\eta^2 + \dd\chi^2 + \sin^2\chi \dd\Omega_2^2\right),
\end{equation}
with the dimensionless conformal time $\eta \in (-\pi, 0)$ and the comoving radial coordinate
$\chi \in [0, \pi]$. Here $H^2 = \Lambda_{\rm eff}/3$.   We use the $(\eta,\chi)$ conformal diagram
throughout to represent the spacetime.

Closed, flat, and open isotropic coordinates can alternately be used to foliate portions of de Sitter
space and are useful in finding solutions to the background massive gravity equations and for investigating perturbations.   With the transformations
\begin{eqnarray}
 \sinh (H t_c) &=& - \cot \eta  ,\nonumber\\
 H r_c &=& 2 \tan(\chi/2),
 \label{eqn:closedtr}
\end{eqnarray}
the line element~\eqref{conformal_metric} takes its closed isotropic form
\begin{equation}
\label{eqn:closeddS}
\dd s^2 = -\dd t_c^2  +\left[ \frac{ \cosh{(H t_c)} }{1 +(H r_c)^2/4} \right]^2 \left(\dd r_c^2 + r_c^2 \dd\Omega_2^2\right),
\end{equation}
where $t_c \in (-\infty,\infty)$, $r_c \in [0,\infty)$. These coordinates chart the entire de Sitter spacetime.
Similarly, defining the coordinates
\begin{eqnarray}
 e^{H t_f} &=&  -\frac{\cos\chi+\cos\eta}{\sin\eta}  ,\nonumber\\
 H r_f &=& \frac{\sin\chi}{\cos\chi+\cos\eta},
 \label{eqn:flattr}
\end{eqnarray}
obtains the flat isotropic form 
\begin{equation}
\dd s^2 = -\dd t_f^2  +e^{2 H t_f}\left(\dd r_f^2 + r_f^2 \dd\Omega_2^2\right),
\end{equation}
where $H t_f \in(-\infty,\infty)$, $H r_f \in [0,\infty)$. These coordinates chart the upper left half of the
conformal diagram $\eta > \chi-\pi$.
Finally, the coordinate definition
\begin{eqnarray}
\ln\left[ \tanh(H t_o/2 ) \right] &=& \tanh^{-1}\left( \frac{\sin\eta}{\cos\chi} \right),\nonumber\\
2 \tanh^{-1}(H r_o/2)&=&\tanh^{-1}\left( \frac{\sin\chi}{\cos\eta} \right),
\label{eqn:opentr}
\end{eqnarray}
gives the open isotropic form
\begin{equation}
\dd s^2 = -\dd t_o^2  +\left[ \frac{ \sinh{(H t_o)} }{1 - (H r_o)^2/4} \right]^2 \left( \dd r_o^2 + r_o^2 \dd\Omega_2^2\right),
\end{equation}
where $H t_o \in (0,\infty)$, $H r_o \in [0,2)$. These coordinates chart the upper left wedge of the conformal diagram 
$\eta > \chi-\pi/2$, corresponding to $1/8$ of the space.   We display these charts with the curves of constant $t_i$ and $r_i$ in Fig.~\ref{fig:coverings}.

\begin{figure}
\center
\includegraphics[width = 0.40\textwidth]{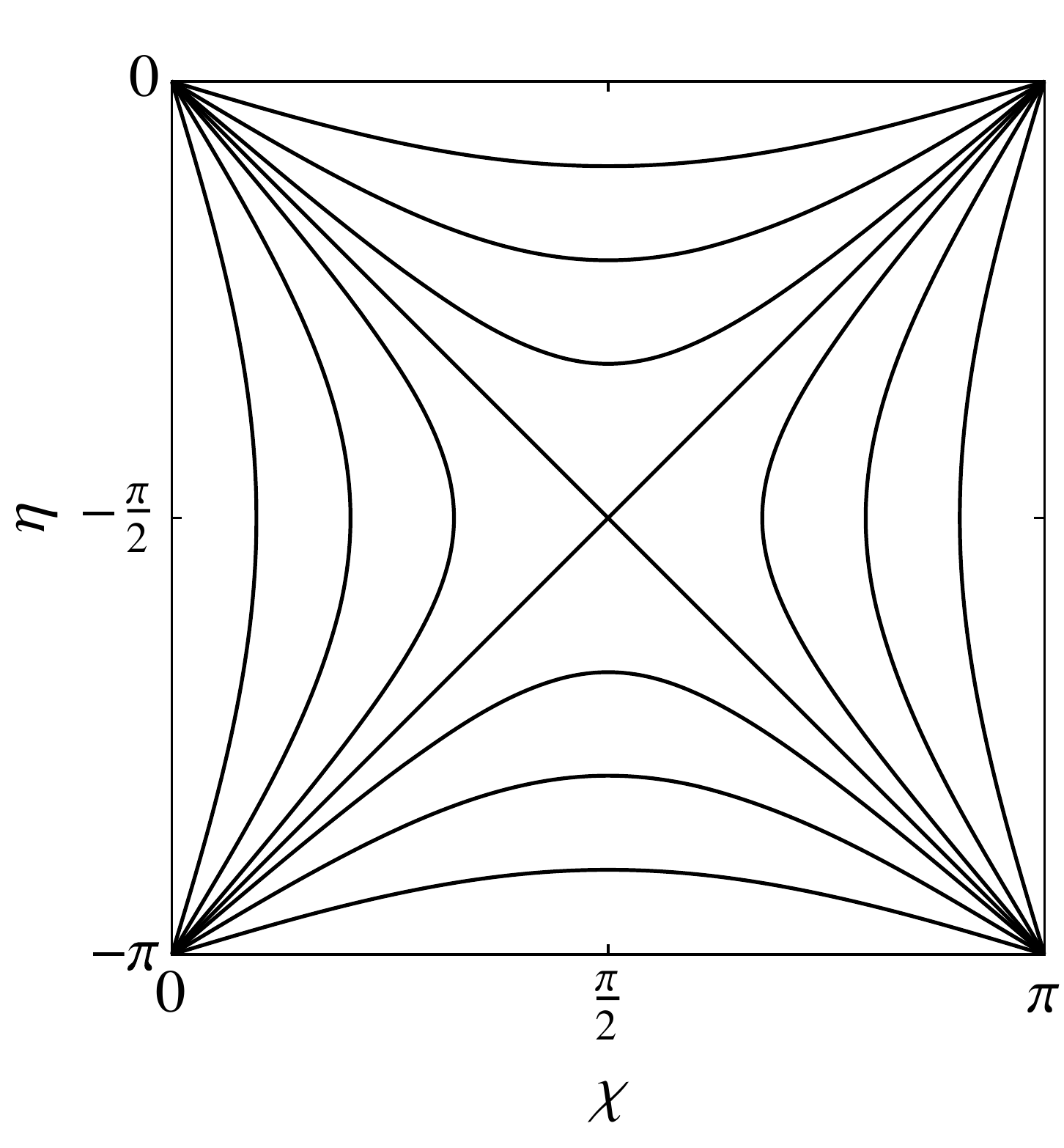}
\caption{Constant unitary gauge radius $g$ curves, {given by Eq.~\eqref{eq:gfordS}}. These are common to all
bi-isotropic solutions.}
\label{fig:const_g}
\end{figure}

Self-accelerating background solutions appear superficially different in different isotropic coordinates.   
More importantly, differences in the constant time surfaces {cause the kinetic terms 
for the perturbations to appear differently in} each case, as emphasized by
Ref.~\cite{Khosravi:2013axa}.
Of course, the causal structure of the
conformal diagram remains unchanged and serves to highlight the spacelike, timelike or lightlike
nature of curves rather than coordinate dependent definitions of simultaneity. For this reason, it will be convenient to plot characteristics on the conformal diagram.

\begin{figure*}
\center
\includegraphics[width = 0.99\textwidth]{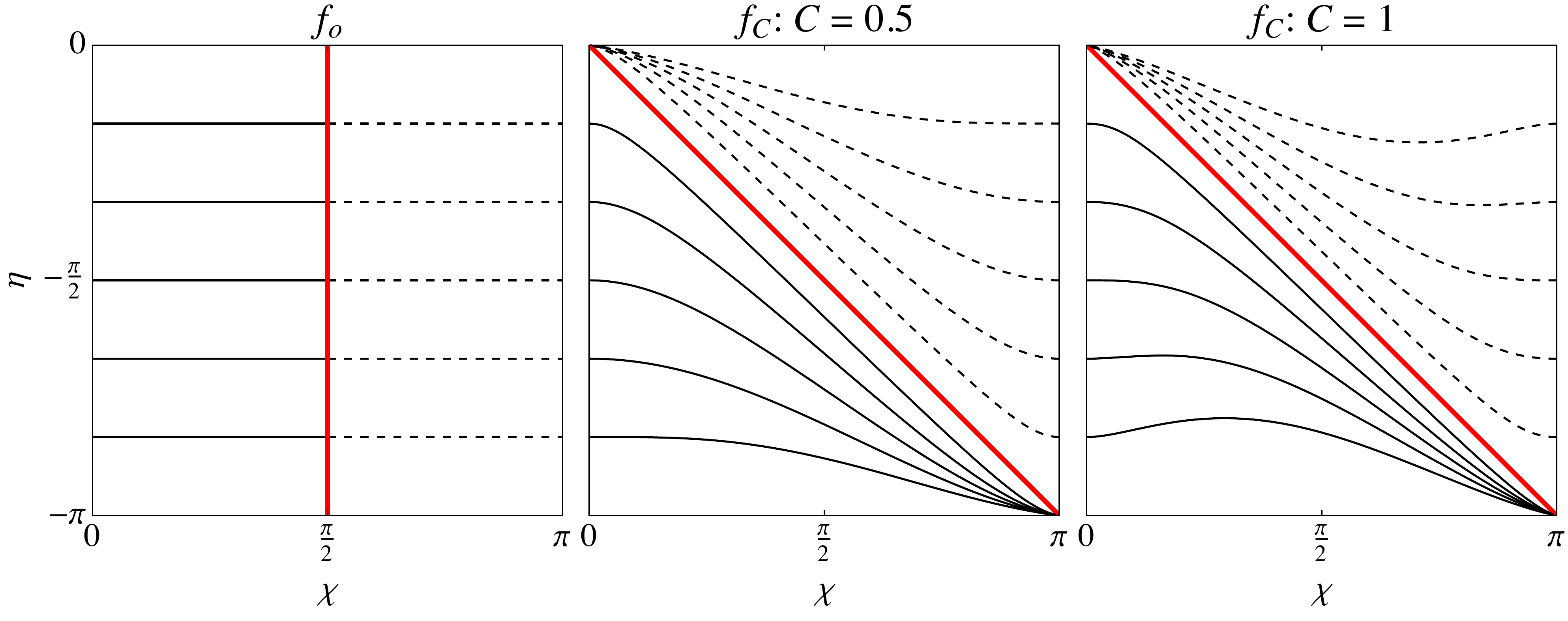}
\caption{Constant unitary time curves for various well-known self-accelerating solutions: open $f_o$ (left), and stationary $f_C$ for $C = 1/2$
(middle) and $f_C$ for $C = 1$ (right).   Solid lines indicate {constant unitary time slices extending from} $\chi=0$ which end at a determinant singularity (red lines).   Dashed lines indicate an extension of solutions with a second copy of the fiducial metric.   }
\label{fig:constant_f}
\end{figure*}

\subsection{Explicit Solutions}
\label{sec:backgroundsolns}

We now combine  the isotropic de Sitter charts with the formalism of \S \ref{sec:backgroundeoms} to construct vacuum solutions to the massive gravity equations of motion.

\vspace{.07cm}
\noindent
{\bf Radial solution: $g$ and equations of motion}

Since closed isotropic coordinates chart the whole spacetime, it is convenient to use these coordinates
to find self-accelerating solutions.  Using the closed line element Eq.~\eqref{eqn:closeddS} in the
radial unitary gauge solution  of  Eq.~\eqref{eqn:gsoln}, we obtain
\be
\label{eq:gfordS}
g = x_0 \frac{  \cosh{(H t_c)} }{1 +(H r_c)^2/4}r_c = -\frac{ x_0}{H} \frac{\sin\chi}{ \sin \eta} .
\ee
Fig.~\ref{fig:const_g} shows the contours of constant $g$ in the conformal diagram. As  the conformal diagram 
makes obvious,
 $g$ is 4-fold symmetric  in the de Sitter spacetime since $\eta \rightarrow -\pi -\eta$ and/or
$\chi \rightarrow \pi -\chi$ provide the same values.

We then look for a choice of unitary time $f$ that solves  Eq.~\eqref{fEOMback}, which in the coordinates~\eqref{conformal_metric} takes the explicit form
\begin{eqnarray}
\label{fEOMbackConformal}
&&
\left( f_{,\chi}^2\cot^2\chi  -
f_{,\eta}^2 \right)\sin^2\eta
+ f_{,\chi}^2 + f_{,\chi} f_{,\eta} \cot\chi\sin(2\eta)  \nonumber\\
&&\quad
= -\frac{x_0^2}{H^2} \csc^2\eta.
\end{eqnarray}
Below we consider three explicit families of solutions to Eq.~\eqref{fEOMbackConformal}, corresponding to different self-accelerating backgrounds whose perturbations we will examine in the subsequent sections.

As an aside, we note that~\eqref{fEOMbackConformal} can be cast in a particularly simple form, 
\begin{equation}
\label{eq:volkovfeq}
 f_{,\tau}^2 - f_{,\rho}^2 =\frac{x_0^2}{H^2},
\end{equation}
 by a further coordinate transform~\cite{Mazuet:2015pea}
\begin{equation}
\rho= \frac{\cos\chi}{\sin\eta}, \quad \tau=-\cot\eta,
\end{equation}
with $\tau \in (-\infty,\infty), \rho \in (-\infty,\infty)$.

\vspace{.07cm}
\noindent
{\bf Open solution: $f_o$}

The simple ans\"atz that $f$ is a function of $\eta$ alone leads to the solutions first
identified in Ref.~\cite{Gumrukcuoglu:2011ew} through an open slicing construction, where $f$ is given by
\be
	\label{solMukohyama}
	f_o(\eta,\chi) 
	= - \frac{x_0}{H} \cot \eta .
\ee
In open slicing, the spacetime and fiducial metrics are bi-diagonal (because Eq.~\eqref{eqn:bidiagonal} is satisfied) and manifestly
homogeneous and isotropic. This property of open slicing applies beyond the de Sitter solutions considered here to a general FRW spacetime, giving this solution a special status.
As pointed out by Ref.~\cite{Khosravi:2013axa}, and rediscovered by
Ref.~\cite{Mazuet:2015pea}, bi-diagonality does not hold in the closed or flat slicings of de Sitter.  
However, the ability to define multiple homogeneous and isotropic slicings and threadings of the spacetime
is special to de Sitter. For a general FRW spacetime, this freedom no longer exists since the homogeneity
of an evolving background density picks out a unique time slicing.

Curves of constant unitary gauge time for this solution are plotted  in Fig.~\ref{fig:constant_f}.
Note that the coordinate pair $(f_o,g)$ maps to two different $(\eta,\chi)$ points in the spacetime 
related by $\chi \rightarrow \pi -\chi$ and correspondingly the determinant of the Jacobian transformation to unitary gauge is zero along $\chi=\pi/2$.
Therefore, more than one copy of the Minkowski fiducial space is required to cover the entire spacetime.
For this solution, the singularity lies in the past light cone of an observer at $\chi=0$ and
hence boundary conditions that represent the continuation with a second fiducial metric
are required. Note that---as mentioned in \S \ref{sec:theory}---such a rule for continuation is  {\it ad hoc} as it must be imposed by hand.

\vspace{.07cm}
\noindent
{\bf Stationary solution: $f_C$}

Self-accelerating solutions where the spacetime metric in unitary gauge is stationary were first identified
in Ref.~\cite{Koyama:2011yg}.   In conformal coordinates, unitary time takes the form
\ba
	\label{solKoyama}
	f_C &=& \frac{x_0}{C H} \left(    \ln \left\lvert \frac{C^2 e^{H t_f}}{1-y}\right\rvert - y \right),\nonumber\\
	y &=& \sqrt{1 + C^2( \sin^2\chi/\sin^2\eta -1)},
\ea
where $C \in (0, 1]$ is a free parameter, $y \in [0,\infty)$, and $t_f(\chi,\eta)$ is the extension of flat isotropic
time to the full de Sitter space using Eq.~\eqref{eqn:flattr}.   The coordinate pair $(f_C,g)$ maps to two spacetime points $(\eta,\chi)$ and
$(-\pi-\eta,\pi-\chi)$.  The inversion singularity thus is at $\chi = -\eta$ and is characterized
by a divergence in unitary time $f_C \rightarrow \infty$ which prevents its differentiability.   Again, two copies of
the Minkowski fiducial space are required to cover the spacetime.  In this case,
the singularity is along the past light cone of the observer at $\chi=0$ at the terminal
conformal time $\eta=0$. Conformal diagrams showing curves of constant $f_C$ for $C = 1/2$
and $C = 1$ are plotted in Fig.~\ref{fig:constant_f}.

A different construction of this solution that starts in static de Sitter coordinates \cite{Mazuet:2015pea} is in fact equivalent to Eq.~\eqref{solKoyama} after
relating their parameter $q$ to $C$ as $q^2 = 1/C^2 - 1$. The equivalence between the two classes of solutions can also be seen directly in the
construction itself since Ref.~\cite{Koyama:2011yg} showed that stationary unitary coordinates and
static de Sitter coordinates are simply related by a radially dependent offset to the static time coordinate.

\begin{figure*}
\center
\includegraphics[width = 0.99\textwidth]{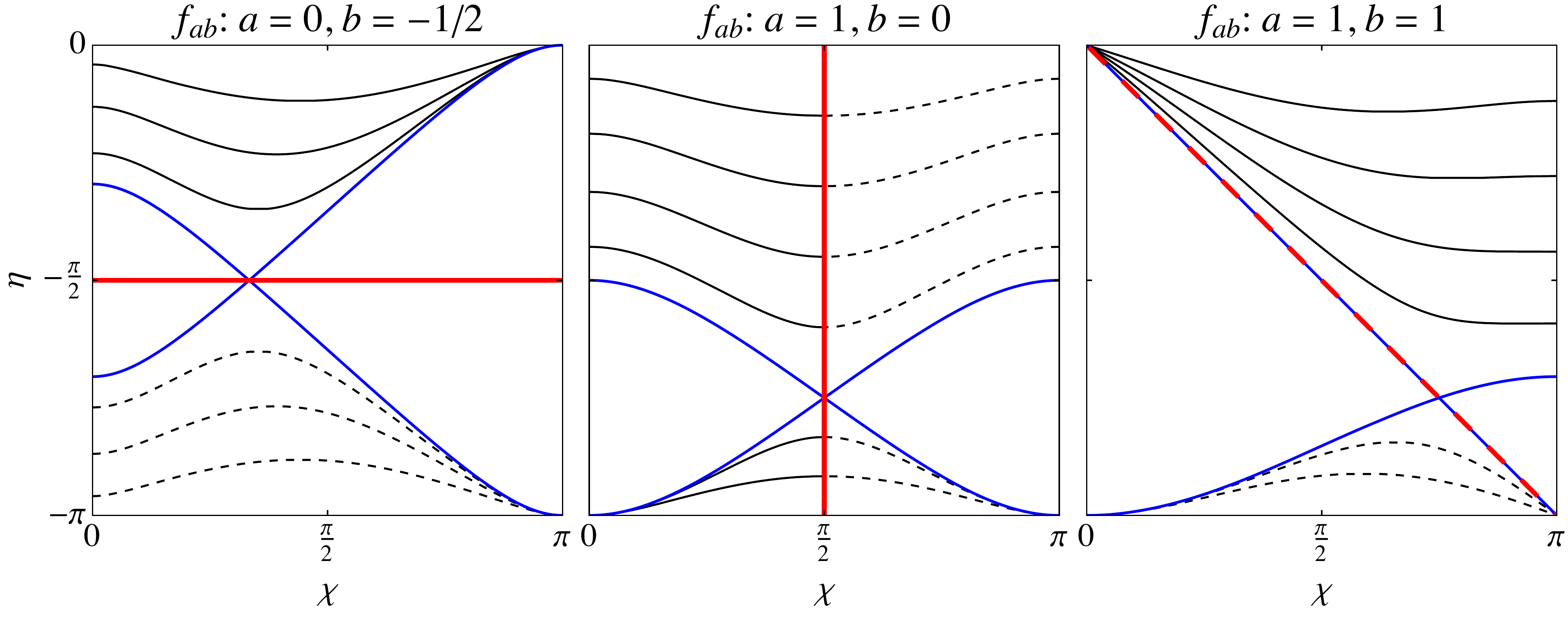}
\caption{Constant unitary time curves for the new $f_{ab}$ class of self-accelerating solutions.
In addition to the zero determinant singularity (thick red lines) there also appear singularities where the determinant is infinite (blue lines), in the $a=b=1$ case the two coincide.  Across the infinite determinant singularity, a na\"ive analytic continuation would require a Riemannian rather than Lorentzian second fiducial metric.}
\label{fig:constant_f_Khosravi}
\end{figure*}

\vspace{.07cm}
\noindent
{\bf New solution: $f_{ab}$}

Finally, we can generalize a particular solution introduced in Ref.~\cite{Khosravi:2013axa}, to a new two
parameter class, where the {temporal} St\"uckelberg field takes the following form
\be
	f_{ab} =\pm \frac{x_0}{H} \sqrt{(a - \cot \eta)^2 - (b - \cos \chi \csc \eta)^2 \vphantom{\big|}}.
\ee
Here, the parameters $a,b$ can take on any real value except for $a=b=0$.   For that case, we have $f_{00}^2 = g^2  - (x_0/H)^2$, such that unitary time and radius cannot identify a unique spacetime point. Ref.~\cite{Khosravi:2013axa} previously considered the case of $a=0$, $b=1$.

In the full spacetime $f_{ab}$ is not guaranteed to be real and so unitary gauge
ends in a determinant singularity where $W = \pm \infty$.   Approaching this point,
the change in unitary time (or proper time measured by the fiducial metric) per unit conformal time  (or proper time measured by the spacetime metric) diverges.
Unlike the
case of $W=0$, an analytic continuation of unitary coordinates
would make a copy of the fiducial metric with a Riemannian rather than Lorentzian signature. 
Thus, unlike the other solutions, it does not appear that $f_{ab}$ can be continued beyond this type of determinant
singularity with copies of the fiducial metric in the same class of solutions.

Changing the sign of $a$ reflects
solutions vertically across $\eta = -\pi/2$ whereas changing the sign of $b$ reflects horizontally across $\chi=\pi/2$.
The  determinant singularities $W =\pm \infty$ occur where $f_{ab}=0$ 
and bound regions past which the solutions cannot be continued within the class.
These singularities intersect at
\begin{equation}
\cot\eta = a, \quad \cos \chi =- \frac{b}{\sqrt{1+a^2}},
\label{eqn:Wcross}
\end{equation}
if $|b| \le \sqrt{1+a^2}$. The $W=0$ singularity 
occurs along the curve $a\cos\chi =b\cos\eta$ and intersects both $W =\pm \infty$
singularities at their crossing point.

We show several examples from this class in Fig.~\ref{fig:constant_f_Khosravi}.  
Displacing $b$ from zero at $a=0$ breaks the $\chi$ or left-right symmetry of the conformal diagram allowing constant unitary time surfaces to foliate the spacetime near one pole
with the other pole hidden behind a $W=\pm \infty$ determinant singularity.
The unbroken $\eta$ or top-bottom symmetry has a corresponding $W=0$ singularity at $\eta=-\pi/2$ across which
solutions can be extended with a second fiducial metric. 
Displacing $a$ from zero at $b=0$ does the converse, allowing constant unitary
time surfaces to foliate the spacetime near both poles at either the top or bottom sections of the diagram with
a $W=0$ singularity at $\chi=\pi/2$, again with a second branch of the solution across it.  
Displacing both equally makes one of the $W =\pm \infty$ curves coincide with
$W=0$, allowing foliation
around  only  one pole
and only on the top or bottom section.

\subsection{Determinant Singularities}

All three families of solutions exhibit determinant singularities. Worldlines of observers can intersect these
singularities
in the bulk of the spacetime, {\it i.e.}, after only a finite amount of  proper time has elapsed.   For other observers, the singularity lies in the past light cone.
Both properties imply that the incompleteness of the solutions there cannot simply be ignored.

In the open $f_o$ solution, a na\"ive 
continuation past the $W=0$ singularity keeps the \stucky\ fields or unitary gauge coordinates continuous and 
differentiable but multivalued.   In the static $f_C$ solution, na\"ive continuation keeps them continuous
but not differentiable at the singularity.    In the new $f_{ab}$ solutions, na\"ive continuation beyond the
$W=\pm \infty$ singularities is not possible and when they intersect at a specific
point in the spacetime bulk they do so at the $W=0$ singularity.

Finally,  by pushing the $W=\pm\infty$ singularities of the $f_{ab}$ case to the top or bottom of the conformal diagram
it can be made to resemble the $f_o$ and $f_C$ solutions.  
The $f_o$ solution picks out the same unitary time surfaces as the limiting case of $a \rightarrow \infty$, $b=0$ since $f_o$ and $f_{ab}$ are then linearly related.  
Likewise, the $a = b\rightarrow \infty$ limit of $f_{ab}$ leads to the same
surfaces of constant unitary time as the
$f_C$ solution for $C \rightarrow 0$ in the upper right portion of the conformal diagram, although in this case $f_{ab}$ is a more general function of $f_C$.  The second
copy of $f_C$ for $C \rightarrow 0$  in the lower left corresponds similarly to $a=b\rightarrow -\infty$.

As we shall see in the next section, the appearance of these singularities 
influences the characteristic curves on which information about perturbations propagate.

\section{Perturbation Characteristics}
\label{sec:perturb}

In this section {we use the method of characteristics to} explicitly solve  the equation of motion \eqref{eqn:gs} for the field fluctuation $\delta\Gamma$ in the three families of background solution studied in \S \ref{sec:backgroundsolns}.    The characteristic curves of this linear differential equation define hypersurfaces along which spherically
symmetric stress-energy perturbations propagate.    The conformal diagram of the curves exemplifies the difference between superluminality, strong coupling, and the well-posedness of the Cauchy problem.

\subsection{Method of Characteristics}

We can solve the $\delta\Gamma$ equation of motion for perturbations propagating on the background solutions using the method of characteristics.
The characteristic curves of this equation also clarify the causal structure of solutions.
Transforming Eq.~\eqref{eqn:gs} from isotropic coordinates to
the conformal coordinates $(\eta, \chi)$ leads to a differential equation of the form
\be
	\label{eqn:gs2}
	 V_\eta \Gs_{,\eta} + V_\chi \Gs_{,\chi} + A \Gs =0.
\ee
Here,
\ba
	V_\eta &=&  f_{,\chi} - \tan \chi \tan \eta\,  f_{,\eta},\nonumber\\
	V_\chi &=&  f_{,\eta} + \(\cot \chi + \csc^2 \eta \tan \chi\)\tan \eta\,
	f_{,\chi} .
	\label{eqn:Vcharacteristic}
\ea
This equation depends on the particular background $f(\eta, \chi)$ around 
which we perturb and describes propagation of the $\Gs$ perturbations in
the regions of de Sitter space where $f$ is defined. 
While $A(\eta,\chi)$ is not important for the construction of characteristics themselves, it does enter
{into determining the field profile for $\delta\Gamma$ along the characteristics.}   For completeness, it is given by
\begin{eqnarray}
\frac{A}{N} &=&  \sin\eta \left[\csc\eta \left(\frac{V_\eta}{N}-R_\eta\right)\right]_{,\eta} \nonumber\\
&& +\cos^2\frac{\chi}{2}\left[ \sec^2\frac{\chi}{2}
 \left(\frac{V_\chi}{N}-R_\chi\right) \right]_{,\chi},
\end{eqnarray}
where
\begin{eqnarray}
R_\eta &=&\frac{\mu}{2} \cos\chi \sin^2\chi \cot\frac{\chi}{2} \csc\eta, \nonumber\\
R_\chi &=&\mu \cos^2\frac{\chi}{2} \sin^2\chi \cot\eta \csc\eta ,\nonumber\\
N&=& \csc^2\chi \sec\chi \tan\eta \Big[  2 \mu\cos\eta\tan\frac{\chi}{2} f_{,\chi}\nonumber\\
&&+ \sec^2\frac{\chi}{2} \left( \mu \cos\chi\sin\eta f_{,\eta} -\frac{x_0}{H} \right)\Big].
\end{eqnarray}

\begin{figure*}
\center
\includegraphics[width = 0.99\textwidth]{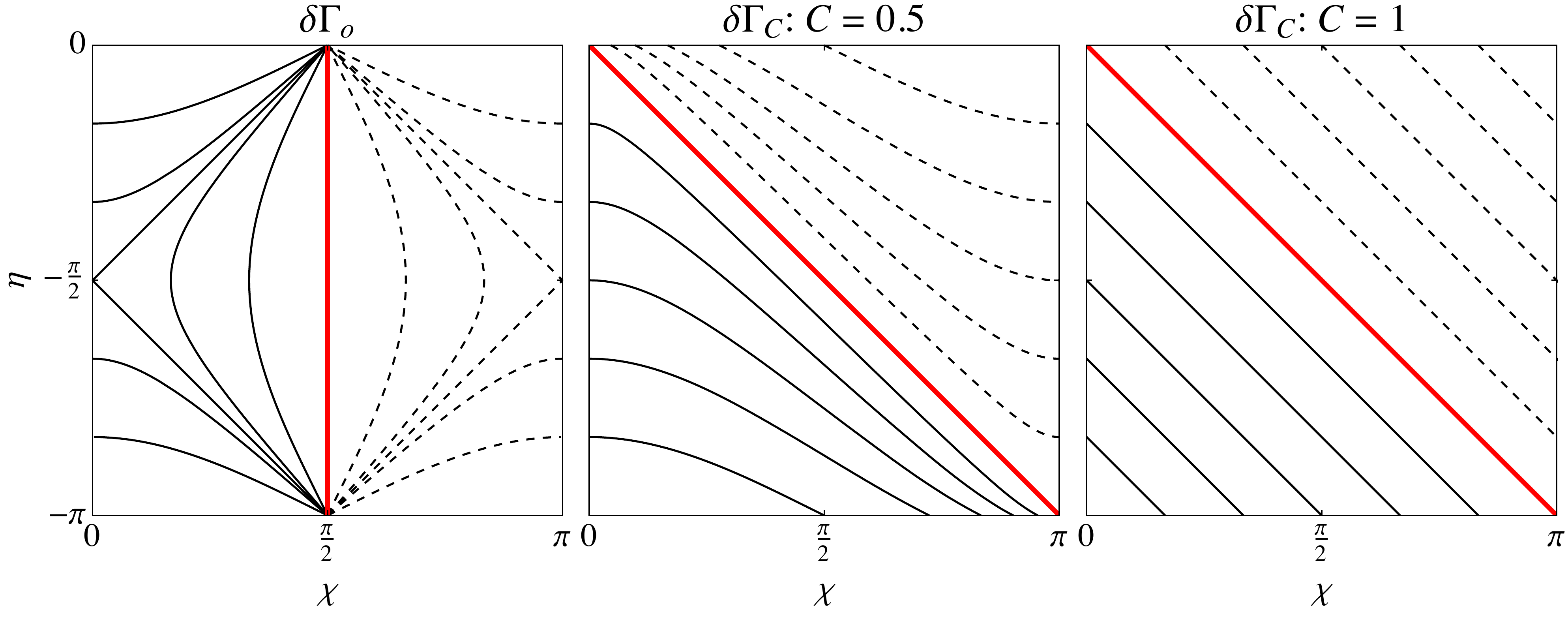}
\caption{Characteristic curves of $\delta \Gamma$ for the 
 background solutions shown in Fig.~\ref{fig:constant_f}.   For $\delta \Gamma_o$,
characteristics coincide with constant open time curves in the open  wedge of 
Fig.~\ref{fig:coverings} and no spacelike surface intersects all characteristics.
For $\delta \Gamma_{1/2}$, characteristics are also spacelike but those within the past light cone at
$\chi=0$ {do all intersect the spacelike surface $\eta = -\pi$.}   For the special case of $\delta\Gamma_1$,
the characteristics are all lightlike.  
Red lines divide the characteristics on either side of the determinant singularity.
 }
\label{fig:characteristics}
\end{figure*}

The most general solution to Eq.~\eqref{eqn:gs2} can be found from its characteristics, which are integral curves for the vector field
\be
	\boldsymbol{V} = (V_\eta, V_\chi) ,
\ee
with the first component in the $\eta$ direction. 
Information from boundary conditions on $\Gs$,  specified on a surface that intersects the characteristic curves, propagates along those curves to provide the general solution in the bulk.
Thus conformal diagrams of
characteristic curves for particular background solutions  present a succinct way of describing
their casual structure.
In particular, the characteristic curves are
\begin{align}
{\rm timelike:} &\quad |V_\eta| > |V_\chi| , \nonumber\\
{\rm lightlike:}  &\quad |V_\eta| = |V_\chi| ,\nonumber\\
{\rm spacelike:}&\quad   |V_\eta| < |V_\chi| .
\end{align} 
Since Eq.~(\ref{eqn:gs2}) is a first order differential equation, the characteristic
analysis yields complete solutions for $\Gs$, including both smooth and discontinuous
cases.   This should be contrasted with a characteristic analysis of a second order
differential equation where in most cases the characteristics only describe the propagation
of discontinuous fronts.
Once a solution for $\Gs$ is specified from boundary data in the conformal coordinate system it can be expressed
in any coordinate system since it transforms  as a scalar
function in spacetime.

\subsection{Explicit  Solutions}

We now apply this formalism to the explicit background solutions derived in \S\ref{sec:backgroundsolns} and identify the characteristic curves upon which isotropic stress-energy perturbations propagate.

\vspace{.07cm}
\noindent
{\bf Open solution: $f_o$}

For the $f_o$ background solution 
\eqref{solMukohyama},  the general solution given by the characteristics is
\ba
	\Gs_o(\eta, \chi) &=& \frac{\sin^2 \eta}{ \sin^2 \chi} F(\phi_o) , \nonumber\\
	\phi_o(\eta,\chi) &=& \cos\chi/\sin\eta ,
	\label{eqn:charopen}
\ea
where $F$ is a completely arbitrary function of its argument $\phi_o$.
Curves of $\phi_o=$\,const., shown in 
Fig.~\ref{fig:characteristics} (left), are the 
characteristics of the equation and comparison with Eq.~\eqref{eqn:opentr} shows that they 
coincide with constant open time $t_o=$\,const.\ surfaces in the wedge charted by the open coordinates.     Correspondingly,
in open isotropic coordinates,  Eq.~\eqref{eqn:gs} loses its kinetic term $(A_{t_o}=0)$ indicating
that initial conditions at $t_o=$\,const.\ do not propagate off that surface.  Open time
slices therefore are disconnected, leading to an unspecified evolution of $\delta \Gamma$. {In particular, information propagates instantaneously along the characteristics as measured by open time.}
This behavior does not violate local conservation of energy and momentum since they are both zero for
any $\delta \Gamma_o$ in these coordinates  \cite{Wyman:2012iw}.

In flat or closed coordinates, these features take a superficially different form.   Characteristics are still spacelike but the finite stress
components in open coordinates transform to energy and momentum.   This energy-momentum is conserved by
the solution \eqref{eqn:charopen} as a consequence of Eq.~\eqref{eqn:gs}. 
As discussed in \S \ref{sec:isopert}, this conservation law is the transformation of the
constraint on the spatial stresses in open coordinates.

On the other hand,
the stress-energy emanates from $\chi=0$ where the spatial profiles of the $\delta\Gamma_o$ 
solutions diverge. Although it is not manifest in the linearized treatment from the quadratic Lagrangian, generically one might expect perturbation theory  to break down here with higher order interactions making the fluctuations strongly coupled.
This class of solution was excluded by the analysis of
Ref.~\cite{Gumrukcuoglu:2011zh} by implicitly demanding regularity at $\chi=0$, leading to their conclusion that the \stucky\ fields 
contained no degrees of freedom.   Our characteristic analysis makes this assumption explicit:
it corresponds to assigning boundary conditions on the timelike $\chi=0$ curve.
More generally, there is no spacelike Cauchy surface at all in the open wedge, since a surface must be timelike to intersect all characteristics.   The lack of a spacelike Cauchy surface is diffeomorphism invariant.  
Thus, even in flat or closed slicing, counting degrees of freedom according to initial data would lead to the
conclusion that the $f_o$ case does not possess a degree of freedom that admits a well-posed Cauchy problem.
The spacetime defined by the solution $f_o$ is an example of a spacetime that is not globally hyperbolic.\footnote{The existence of such a pathological spacetime as a solution to dRGT does not imply pathologies inherent to the theory itself; indeed, general relativity admits solutions which are not globally hyperbolic.}

While imposing a regular boundary  at $\chi=0$ for these characteristics might seem reasonable in the open wedge itself, 
it is interesting to track characteristics intersecting this boundary in different parts of the de Sitter space.  Fig.~\ref{fig:characteristics} (left) shows that in the lower left wedge, where the characteristics are mirror
images of the open wedge, there is a spacelike Cauchy surface at $\eta\rightarrow -\pi$ despite 
superluminality, but the characteristics end rather than begin at $\chi=0$.   Stress-energy on
a characteristic then flows into the origin requiring a boundary condition or nonlinear completion of the theory to specify its further evolution.  Thus non-singular behavior on the open wedge at $\chi=0$ may require special initial conditions in the lower left wedge.  On the other hand this behavior is analogous to collapse of a 
perfectly spherically symmetric density shell in ordinary linearized, Newtonian gravity.  The finite angular momentum of generic perturbations may prevent singular behavior---we leave investigation of this possibility
to future work.    In any case, while the inwardly directed characteristics might signal strong-coupling at $\chi=0$ where the fluctuations diverge, like a black-hole nothing emanates classically from this point within the lower left wedge.

Interestingly, in the central diamond the $\delta\Gamma_o$ characteristics 
are timelike, or subluminal, and admit a well-posed Cauchy problem.    However, the characteristics
are split by the determinant singularity into  left and right halves.   Although information in
$\delta\Gamma_o$ never crosses this singularity in the linearized theory, the worldlines of other particles can.  From the perspective of a massive gravity theory with only one copy of the
fiducial metric, the determinant singularity appears as another spatial boundary condition.  Even allowing the theory to be extended, the boundary condition must be imposed by hand at each order in perturbation theory, {\it e.g.}, by demanding that 
the \stucky\ fields are smooth across the boundary to the second copy of the fiducial metric
\cite{Gratia:2013gka}.

\begin{figure*}
\center
\includegraphics[width = 0.99\textwidth]{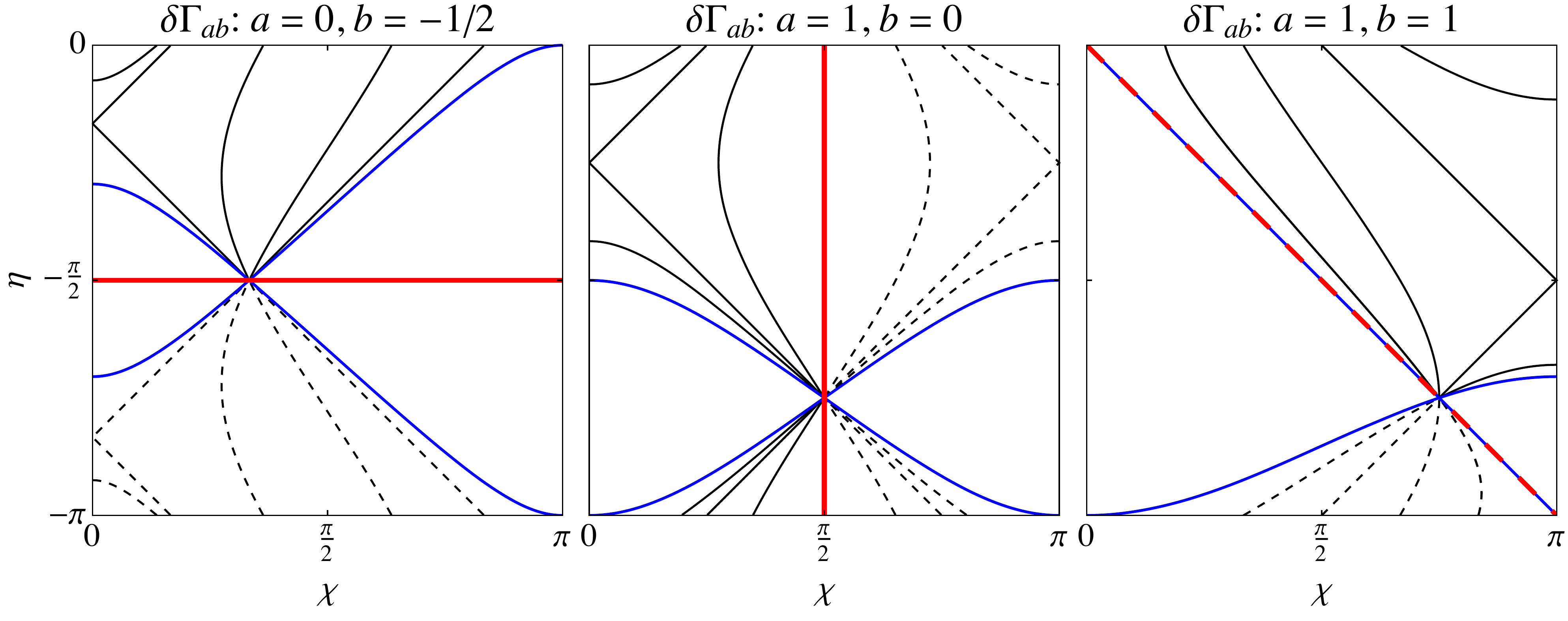}
\caption{Characteristic curves of $\delta \Gamma_{ab}$ for the new class of
$f_{ab}$ solutions illustrated in Fig.~\ref{fig:constant_f_Khosravi}.   The determinant singularities
strongly influence the causal structure of the characteristics. These backgrounds share many of the features of the $f_o$ and $f_C$ spacetimes discussed in Fig.~\ref{fig:characteristics}
and provide new ones due to the $W=\pm \infty$ determinant singularities (blue curves).
}
\label{fig:characteristics_Khosravi}
\end{figure*}

\vspace{.07cm}
\noindent
{\bf Stationary solution: $f_C$}

The characteristics of perturbations around the $f_C$ backgrounds share some, but not all, of the properties of the $f_o$ background.
Here Eq.~\eqref{eqn:gs2} is solved by
\ba
	\Gs_C(\eta, \chi)& = &
	\frac{ \sin^2 \eta}
		{ \sin^2\chi } \frac{F(\phi_C)}{y(\eta,\chi)} ,\nonumber\\
	\phi_C(\eta,\chi)&=& 
 	\frac{\sin \eta [1-y(\eta,\chi)]}
		{\cos \chi + \cos \eta},
\ea
where $F$ is again an arbitrary function of its argument and $y$ was defined in 
Eq.~\eqref{solKoyama}.
In Fig.~\ref{fig:characteristics} (mid), we show the $\phi_C=$ const.\ curves for $C=1/2$.
These characteristics are spacelike across the whole spacetime. 
Even though $\phi_C=$\,const.\ does not coincide with the open, flat or closed choice of
isotropic time, these surfaces foliate the spacetime.   Thus in principle we could define
a new choice of time $t_C = \phi_C$ for which the kinetic term for $\delta \Gamma_C$
vanishes, just like it does for $\delta \Gamma_o$ in the open slicing.  With this choice of time coordinate, information 
along the characteristics again propagates instantaneously and energy-momentum conservation becomes a spatial constraint.

On the lower left half of the $\delta\Gamma_C$ diagram, the causal structure resembles the lower left wedge of the $\delta \Gamma_o$ case.
Here $\eta \rightarrow -\pi$ provides a spacelike Cauchy surface and the characteristics
end at $\chi=0$ with a divergent $\delta \Gamma_{1/2}$ radial profile.   On the upper
right half, the structure resembles the open wedge of $\delta \Gamma_o$ with characteristics
emanating from the other pole of the closed de Sitter space.   However, the two halves
are divided by the null line that designates the determinant singularity where $f$ is not
continuously differentiable.    Thus the outgoing characteristics of upper right half are not within the past light cone at $\chi=0$ unlike
the $\delta \Gamma_o$ case.   This example illustrates that superluminality, a well-defined
Cauchy problem, and singular field profiles in the past that might indicate problems originating from a point of strong coupling are not one-to-one related.

In the special case of $C=1$, $y=-\sin\chi/\sin\eta$  and the characteristics of $\delta\Gamma_1$
become null (see Fig.~\ref{fig:characteristics} and Ref.~\cite{Wyman:2012iw}).    This special case was interpreted by
Ref.~\cite{Koyama:2011yg} as having no vector in the background, based on a
decoupling limit analysis.   Although this choice
eliminates the superluminality found in other solutions, the lightlike determinant singularity still 
exists for this case.   While it 
only occurs on 
the past lightcone of the observer at $\chi=0$, $\eta=0$, it would be within the past light
cone at other points in the spacetime and still require null boundary conditions to define
global solutions.  Likewise other observers in the lower left region would reach the determinant
singularity after finite proper time. Even with solutions continued beyond the singularity these observers would begin
to cross characteristics for which no spacelike Cauchy surface exists.

\vspace{.07cm}
\noindent
{\bf New solution: $f_{ab}$}

For the new class of $f_{ab}$ backgrounds, the general solution is
\ba
	\Gs_{ab}(\eta, \chi)& = &
	\frac{\csc^2 \chi \sin^3 \eta}{\cos \eta - a \sin
	\eta}F(\phi_{ab}) ,\nonumber\\
	\phi_{ab}(\eta,\chi)&=& 
 	\frac{\cos \chi - b \sin \eta}
		{\cos \eta - a \sin \eta} .
\ea
This solution serves to illustrate that the nature of the characteristics is strongly influenced 
by the global structure of the determinant singularities discussed in \S\ref{sec:backgroundsolns}.  
Characteristics run tangent to these singularities as information in the perturbations cannot 
cross these curves (see Fig.~\ref{fig:characteristics_Khosravi}).  {If $|b| <
\sqrt{1+a^2}$, the $W=\pm\infty$ and $W=0$
singularities intersect within the spacetime bulk at a point given by
Eq.~\eqref{eqn:Wcross}.   
Though the characteristics tangent to these singularities} also intersect, 
the perturbation amplitude $\Gs_{ab}$ diverges at this point and hence marks a point where strong
coupling is likely.

Null lines that emanate from this
point serve as separatrices which divide regions of super- and subluminal propagation much like the
$f_o$ case.    This family of examples
shows that these null lines are not necessarily related to special curves in the
open, closed or flat de Sitter slicing.    Where these null lines intersect the poles, the characteristics
change from ingoing into the pole, which can be all be intersected by a single 
spacelike Cauchy surface in the past,  to
outgoing which cannot.   For the special case that $|a|=|b|$ one of these null lines
coincides with the determinant singularity.

\subsection{Superluminality and Bi-Isotropy}

In all but one example {($\delta\Gamma_C$ with $C=1$)},  the characteristics 
intersect the poles $\chi
= 0, \pi$ at right angles {at all but a finite number of points}.  
In conformal coordinates, perturbations propagate 
instantaneously there.
As we now show, this is a generic property of bi-isotropic solutions. In our explanation we focus only on the $\chi = 0$
pole as treatment of the other pole is analogous. Expanding  unitary time around
$\chi = 0$ in a series
\be
\label{fExpansion}
	f(\eta,\chi) =  h_0(\eta) + h_1(\eta) \chi + h_2(\eta) \chi^2 + {\cal{O}}(\chi^3)
\ee
and using the $\chi$ expansion of the equation of motion \eqref{fEOMbackConformal}, 
one can show that $h_1$ is always identically zero. Physically this means that 
the bi-isotropy condition and the 
ability to make arbitrary Lorentz boosts in the fiducial space ensures that unitary time
and conformal time can be locally aligned.    The difference between surfaces of
unitary time and conformal time thus  grow at most quadratically with the distance to the
pole $\chi^2$ or more generally differ only by the curvature corrections to a locally 
flat spacetime metric.  Since the same is true between conformal time and closed, flat, or
open isotropic time, instantaneous propagation in conformal coordinates implies
instantaneous propagation in isotropic coordinates at the poles.  In fact, the
arguments below for the generality of spacelike characteristics
apply beyond the vacuum cases considered here to spacetime metrics
that are  locally flat near the bi-isotropy point.

Expanding the vectors defined by Eq.~\eqref{eqn:Vcharacteristic} in the same fashion,
\begin{equation}
\boldsymbol{V} = \left(
\begin{array}{c}
 {\cal O}(\chi)  \\
  2  h_2 \tan \eta + h_{0,\eta} + {\cal O}(\chi)  \\
\end{array} 
\right) .
\end{equation}
In a neighborhood of a general point $(\eta_0, 0)$ the leading order terms are
finite  and the vector field $\boldsymbol{V}$ is aligned with the curves of  
constant conformal time $\eta$, signaling instantaneous propagation. Using the expansion of the background equation of motion
\eqref{fEOMbackConformal}, it is possible to relate
$h_2$ and $h_{0,\eta}$ to prove that infinite superluminality at the given point
is avoided if and only if
\be
\label{LuminalBC}
	h_{0,\eta}(\eta_0) = \pm \frac{x_0}{H \sin \eta_0} .
\ee
As seen with our examples, this condition can be satisfied for all $\eta$ ($\Gs_C$ with $C =
1$) or at a discrete set of points ($\Gs_o$ and typical $\Gs_{ab}$).

In fact, 
for the special points $\eta_0$  at which this condition is satisfied, the 
characteristics are always luminal, never subluminal.   Expanding $f$ to third order in $\chi$ allows $\boldsymbol{V}$ to be expanded into linear order in
both $\chi$ and $\eta - \eta_0$.  With repeated use of the equation of motion
\eqref{fEOMbackConformal} we then find
that each characteristic curve hitting $\chi = 0$ at $\eta_0$ is luminal.
For $h_3(\eta_0)=0$ there are two such characteristic curves---one incoming and one outgoing---forming the typical luminal separatrices
we see in Fig.~\ref{fig:characteristics} (left) and \ref{fig:characteristics_Khosravi}.
Even for these cases, the characteristics are superluminal at all but a finite
number of points.
 For 
$h_3(\eta_0)\ne 0$, there is only one characteristic which is either incoming or outgoing,
depending on the sign of  $h_{0,\eta}/h_3$.  The $C=1$ example of
Fig.~\ref{fig:characteristics} (right)  exhibits a unitary time solution $f_1$ with these
very special properties.   Together with solutions trivially related by $f \rightarrow -f$ and/or $\eta \rightarrow -\pi
- \eta$, it is in fact the unique bi-isotropic vacuum solution that evades superluminality
entirely as can be shown by integrating the equation of motion \eqref{fEOMbackConformal}
from the boundary conditions \eqref{LuminalBC} along null coordinates.

\section{Discussion}
\label{sec:discuss}

We have investigated the causal structure of three families of vacuum solutions in dRGT massive gravity. 
In particular, we have constructed the conformal diagram of characteristic hypersurfaces
for isotropic stress-energy perturbations around these backgrounds by exploiting the first-order structure of their equation of motion. 
These examples  provide fertile ground for studying the interplay between superluminality, an ill-posed Cauchy problem, and strong coupling as well as issues that arise for the global structure of spacetime in a theory with two metrics.

The $f_o$  solution of~Ref.~\cite{Gumrukcuoglu:2011ew} manifests aspects of all of these issues.
This solution is distinguished because in open slicing both the background spacetime and fiducial metric are simultaneously homogeneous and isotropic. 
In {this} slicing, the kinetic term of perturbations vanishes, 
which was taken as an indication of possible problems with strong coupling \cite{Gumrukcuoglu:2011zh}; this is supported by instabilities identified around anisotropic backgrounds~\cite{DeFelice:2012mx}, and the lack of an isotropic degree of freedom identified
by a Hamiltonian analysis \cite{Khosravi:2013axa}.  However, its kinetic term and Hamiltonian degrees of freedom only vanish  in open slicing  \cite{Khosravi:2013axa}.

Our characteristic analysis clarifies these issues. 
Absence of a kinetic term in the quadratic action in a particular
choice of slicing indicates instantaneous propagation of perturbations in the given frame and more generally perturbative superluminality or spacelike characteristics in any frame.  The Hamiltonian
analysis fails in the pathological choice of slicing along characteristics since the Hamiltonian
is no longer associated with time evolution.    Energy-momentum conservation likewise
becomes a spatial constraint equation, confusing the counting of degrees of freedom. 
While it is the easiest of the interrelated issues to diagnose, it does not necessarily indicate
strong coupling nor an ill-posed Cauchy problem.  Through a particularly poor choice of coordinates, it is possible to make the kinetic term for superluminal fluctuations disappear
even in a free theory.  Likewise, information cannot propagate forward from a pathological choice of initial value surface.

On the other hand,
for the $f_o$ solution, the characteristic analysis does {additionally} reveal an ill-posed Cauchy problem.
Since  spacelike characteristics originate from the spatial origin in the open chart of de Sitter,
there is no spacelike
surface that intersects all characteristics.   Hence this particular spacetime does {\it not} admit a well-defined {initial value} problem, as it lacks a spacelike Cauchy surface. 
In contrast to superluminality, nonexistence of spacelike Cauchy surface always implies the 
quadratic action is pathological.    Further, in contrast to the Hamiltonian identification of degrees
of freedom, this concept and its relation to identifying degrees of freedom by the amount 
of initial data required is diffeomorphism invariant.

Finally, for the $f_o$ solution, finite perturbations specified along characteristics diverge in amplitude
at the spatial origin from which they emanate in the open chart.   Given that the theory 
contains nonlinear interactions,  this indicates that perturbation theory almost certainly 
breaks down leaving the theory strongly coupled there.   Hence in this case, the ill-posed
Cauchy problem likely originates from a point of strong coupling.   
More precisely, strong coupling occurs when the effective field theory breaks down due to ever higher order 
interaction terms becoming important.    While strong coupling cannot be diagnosed
from the quadratic action alone, the divergence or discontinuity of perturbative solutions at a given spacetime point is a good indicator of strong coupling there.

Vanishing of kinetic terms is
often used as a related proxy for strong coupling: once variables are canonically normalized, interaction terms pick 
up negative powers of the coefficient in front of the kinetic term, which drives the effective 
strong coupling energy scale to zero when the kinetic term vanishes.  However,
our example shows that without examining the higher order terms themselves, one 
cannot immediately determine whether the vanishing of kinetic terms instead simply arises from a poor
choice of coordinates.

The $f_C$, or stationary class of solutions, serves to further distinguish these concepts.
Here generic solutions also admit spacelike characteristic curves and hence 
superluminality.   By choosing a time coordinate that is orthogonal to these spacelike surfaces, we again recover a representation where kinetic terms and dynamics vanish in favor of spatial 
constraints.   
Unlike the open $f_o$ solution,
all characteristics of this class that are within the past light cone of an observer at the spatial origin intersect a spacelike
surface.   Hence the Cauchy problem is well-posed for this family of solutions and observers.
Furthermore, the characteristics  end rather than begin at the spatial origin
where the field perturbations  diverge.   While strong coupling at the origin is 
again likely, this distinction is crucial.   {As in spherical collapse of perturbations in
general relativity (with Newtonian gravity as the corresponding effective theory), the formation of a singularity at the origin does not necessarily invalidate the
effective theory far from the singularity if nothing escapes it classically.}

Finally, the new class of $f_{ab}$ solutions serves to highlight issues with the 
global structure of the spacetime that can occur in theories with two metrics.   In all three
families of solutions, the fiducial  Minkowski space
does not cover the whole  de Sitter
spacetime.    Specifically, there are points where the diffeomorphism-invariant ratio of the determinants of the physical and fiducial metrics vanishes, diverges or is undefined.   
Here the spacetime becomes geodesically incomplete.   While the $f_o$ and $f_C$ 
solutions allow straightforward but {\it ad hoc} analytic continuation past these
determinant singularities {to define a solution in the whole spacetime},
the $f_{ab}$ ones do not.   Furthermore, since characteristics do not cross these singularities, continuity
must be imposed not just for the background but also for the perturbations.

The superluminality exhibited in all three classes of solutions is a necessary condition of the
bi-isotropic construction except in one unique case.    Generically, starting at the bi-isotropy point characteristics 
are superluminal across separations comparable to the spacetime curvature as a consequence of the alignment 
between
the locally flat spacetime metric with the fiducial flat metric.    Exceptional
cases produce luminal  but never subluminal characteristics that intersect this {worldine}. 
For the single case of the stationary $f_1$ solution, all characteristics are strictly luminal.
Unlike the usual second-order system where  characteristics define 
only a front velocity of discontinuous solutions,  the characteristic analysis of our first-order
system is fully general.  In particular, it allows the construction of smooth wavepackets that also propagate superluminally and do not on their own imply strong-coupling.

While these examples do serve to distinguish the concepts of superluminality, spacelike
Cauchy surfaces and strong coupling, it is important not to overinterpret their consequences
for the dRGT theory itself.  They represent examples of tree-level or classical propagation of
perturbations on specific, perhaps pathological, self-accelerating backgrounds.
{Superluminality in the full theory} would be in tension with the dRGT theory  admitting a local and Lorentz-invariant UV completion~\cite{Adams:2006sv}. We emphasize, however, that the characteristic analysis alone is insufficient to establish whether or not superluminal propagation is truly present---in order to discern this, it has been claimed that one must know about the Green's function at infinite frequency~\cite{Dubovsky:2007ac,Shore:2007um,deRham:2014lqa}, which goes beyond the classical approximation that the characteristic analysis entails.

It is also worth mentioning that the presence of superluminality does not in and of itself imply acausality~\cite{Babichev:2007dw}. Indeed, in order to establish acausality, one would need to use the superluminal propagation to construct a closed timelike curve ({\it e.g.}, by using a point at which characteristic curves cross). Interestingly, we find (at least in the highly symmetric case we consider) that characteristic curves only
intersect at singularities, indicating that such perturbations remain causal within the effective theory, even if we equate perturbative superluminality with true superluminality. It has  been conjectured in Ref.~\cite{Burrage:2011cr} that this might be true in general: the chronology protection conjecture asserts that traversing a closed timelike curve cannot be achieved within  the regime of validity of the effective theory.

Finally in all of our examples, the Hamiltonian associated with isotropic perturbations
is unbounded {\cite{Khosravi:2013axa}}.   While it is not clear how to interpret the Hamiltonian in these cases where
the equations of motion are first order and 
there exists pathological choices of slicing where it is no longer associated with time-evolution,
this may indicate that the features uncovered here are associated with an unstable background rather than generic consequences of the dRGT theory.

\smallskip{\em Acknowledgments.---}
We thank Rachel Rosen and Robert M.\ Wald for useful discussions.  
This work was supported by
the Kavli Institute for Cosmological Physics at the University of
Chicago through grants NSF PHY-0114422 and NSF PHY-0551142.  
PM was additionally supported by grants NSF PHY-1125897 and NSF PHY-1412261 and an endowment from the Kavli
Foundation and its founder Fred Kavli, WH by U.S.~Dept.\ of Energy
contract DE-FG02-13ER41958
and HM by a Japan Society for the Promotion of Science
Postdoctoral Fellowships for Research Abroad. 
AJ was supported in part by the Robert R.\ McCormick Postdoctoral Fellowship.

\bibliography{cauchy}

\end{document}